\documentclass[12pt]{article}

\usepackage{amsmath}
\usepackage{amssymb}
\usepackage{cite}
\usepackage{color}

\def\rmd{{\rm{d}}}

\def\nn{\nonumber}

\numberwithin{equation}{section}

\title{\bf \Large Perfect fluid equations
with $N=1,2$ Schr\"odinger supersymmetry}

\author{Timofei  Snegirev${}^{a}$\thanks{timofei.v.snegirev@tusur.ru}
\\[0.5cm]
\it{\small ${}^a$Laboratory of Applied Mathematics and Theoretical Physics,}\\
\it{\small Tomsk State University of Control Systems and Radioelectronics,}\\
\it{\small Lenin ave. 40, 634050 Tomsk, Russia}}

\date{}

\begin{document}

\maketitle

\begin{abstract}
Superconformal extensions of the perfect fluid equations, which realize
$N=1,2$ Schr\"odinger superalgebra, are
constructed within the Hamiltonian formalism. They are built by
introducing real (for $N=1$) or complex (for $N=2$) anticommuting field
variables as superpartners for the density and velocity of a fluid.
The full set of conserved charges associated with the $N=1,2$
Schr\"odinger superalgebra is constructed. Within the Lagrangian formalism,
when the Clebsch decomposition for the velocity vector field is used, the
anticommuting variables can be interpreted as potentials
parameterizing fluid's vorticity.

\end{abstract}

\thispagestyle{empty}
\newpage
\setcounter{page}{1}

\allowdisplaybreaks

\section{Introduction}\label{Sec1}

In modern theoretical physics, the construction of supersymmetric extensions of a particular
dynamical system is a natural avenue to explore. In the
case of non-relativistic fluid mechanics, which we consider in this
paper, supersymmetric models are of interest for several reasons.
Firstly, systems of this type are tightly connected with the superstring
theory. For example, the
supersymmetric Chaplygin gas equations in $(1+1)$ and $(1+2)$
dimensions follow from the Nambu-Goto action for the superstring and
supermembrane \cite{JP00,BJ01}. Secondly,
various hydrodynamic systems in one spatial dimension
are completely integrable \cite{LL87,LL60,BC66,ON88,ANNOV89}.
Supersymmetrization of such models can provide new non-trivial
examples of integrable systems in field theory \cite{DP02}. Thirdly,
the construction of extended hydrodynamic systems with
supersymmetries is an efficient means to describe fluid dynamics
with spin degrees of freedom (see Section 4 in
\cite{JNPP04}). Related studies in supersymmetric fluid
mechanics, including exact  solutions, can be found in
\cite{FKT89,GH11,GH13,Gal24}.

It is well known that, for a specific equation of state, the
non-relativistic perfect fluid dynamics, which is described by the
continuity equation for the density and the Euler equation for the
velocity vector field, enjoys the Schr\"odinger symmetry group
\cite{HH1,RS00,HZ}. In addition to the Galilei symmetries, it
includes two conformal transformations: dilatation and special
conformal transformation. Although the Schr\"odinger group was
originally introduced as a kinematical symmetry group of the
Schr\"odinger equation for a free massive particle
\cite{Jack72,Nied72,Hag72}\footnote{ In fact, similar
non-relativistic conformal structure has been known since 19th
century due to the works on classical mechanics\cite{Jac1884} and
the heat equation \cite{Lie1881}. The author thanks an anonymous
reviewer for drawing his attention to this fact.}, it has been found
to be relevant for a wide range of physical applications, especially
in condensed matter physics (for a review see \cite{DHHR24} and
references therein). While supersymmetric extensions of the
Schr\"odinger algebra have been known for a long time
\cite{BH86,GGT90} its dynamical realization in the context of fluid
mechanics remains almost completely unexplored. Note that the latter
can be of potential interest from the standpoint of non-relativistic
holography \cite{BM08,MP08,GMP13} and flud/gravity correspondence
\cite{Ran09}.

Recently, equations of a fluid mechanics, which hold
invariant under the $N=1$ Schr\"odinger
supergroup,  were constructed
within the group-theoretic
approach \cite{Gal24}. They were formulated in terms of
superfields in $R^{(1|1)}$ superspace with one Grassmann-odd
coordinate as the superpartner for the temporal variable. In this
simplest option of superspace, spatial Grassmann-even coordinates are
inert with respect to supertranslations. A peculiar feature of the
construction in \cite{Gal24} is the presence of an extra vector
superfield and, as a consequence, the presence of an extra bosonic
vector component in addition to the fluid velocity.

The goal of this work is to propose an alternative (non-superfield)
approach to building fluid equations with $N=1,2$ Schr\"odinger
supersymmetry, which relies upon the Hamiltonian formalism. In the
past, this approach has been successfully applied to derive
supersymmetric Chaplygin gas models in lower dimensions \cite{JP00}.
Technically speaking, the construction of an $N=1$ supersymmetric
extension of a fluid model with the Hamiltonian $H$ amounts to
introducing fermionic field partners for the bosonic degrees of
freedom and constructing a single real supercharge $Q$, which via
the Poisson bracket produces a super-extended Hamiltonian
$\{Q,Q\}=2iH_S$. The latter governs the dynamics of the enlarged
system and reduces to the original boson theory in the bosonic
limit. The $N=2$ case can be considered in a similar way. To the
best of our knowledge, $N=2$ super-Schr\"odinger fluids have not yet
been studied in the literature.

The work is organized as follows. In the next section, we briefly
remind the structure of the $N=1$ and $N=2$ Schr\"odinger
superalgebras. In Section \ref{Sec3}, the non-relativistic perfect
fluid dynamics and its symmetries are reviewed. As is known, the
full symmetry group can be extended to the Schr\"odinger group for a
specific equation of state \cite{RS00}. In Section \ref{Sec4}, real
fermionic superpartners are introduced for the density and the
velocity vector field and a super-extended perfect fluid model is
build which realizes $N=1$ Schr\"odinger supersymmetry within the
Hamiltonian formalism. It is shown that a simple supersymmetry
charge linear in the fermionic variables suffices to produce a
reasonable super-extended Hamiltonian. The energy-momentum tensor
and its superpartner, as well as the complete set of conserved
charges, are given in explicit form. The Lagrangian description
based upon the Clebsch decomposition for the velocity vector field
is discussed. In Section \ref{Sec5}, introducing a pair of complex
conjugate fermionic superpartners for the density and velocity, a
novel dynamical fluid model with $N=2$ Schr\"odinger supersymmetry
is constructed. In contrast to the $N=1$ case, supersymmetry charges
include a cubic contribution in fermionic variables. In Section
\ref{Sec6}, we summarize our results and discuss possible further
developments.

\section{$N=1$ and $N=2$ Schr\"odinger superalgebras}\label{Sec2}

The Lie algebra associated with the Schr\"odinger group
\cite{Nied72} includes the generators of temporal translation $H$,
dilatation $D$ and special conformal transformation $K$, which form
$so(2,1)$ subalgebra, as well as spatial translations $P_i$, the
Galilei boosts $C_i$ and spatial rotations $M_{ij}$, where
$i=1,...,d$ and $d$ is the spatial dimension. They obey the
structure relations
\begin{align}\label{ShrAl}
& {[H,D]}=H,  && [H,C^{}_i]=P_i,\nn
\\
& {[H,K]}=2D,  && [P_i^{},C_j^{}]=\delta_{ij} m,\nn
\\
& {[D,K]}=K, &&
[P_i,M_{jk}^{}]=\delta_{ij}P_{k}-\delta_{ik}P_{j},\nn
\\
& {{[D,P^{}_i]}=-\frac12P^{}_i}, &&
[C_i,M_{jk}^{}]=\delta_{ij}C_{k}-\delta_{ik}C_{j},\nn
\\
& {{[D,C^{}_i]}=\frac12C^{}_i}, &&
[M_{ij}^{},M_{ab}^{}]=\delta_{i[a}M_{b]j}-\delta_{j[a}M_{b]i},\nn
\\
& {{[K,P^{}_i]}=-C^{}_i},
\end{align}
where the second column defines the Galilei subalgebra with a
central charge $m$. The Schr\"odinger algebra (\ref{ShrAl}) is
sometimes presented in compact notation which is convenient for
constructing infinite-dimensional extensions. For readers's
convenience it is displayed in Appendix.

An $N=1$ supersymmetric extension of the Schr\"odinger algebra, which was first
revealed when studying the model of a non-relativistic spin-$\frac12$ particle
\cite{GGT90}, involves additional Grassmann-odd generators which include the
supercharge $Q$, the generator of superconformal transformation $S$, and the
superpartners $\Lambda_i, i=1,...,d$ of the Galilei boosts. All together they obey the
following (anti)-commutation relations \cite{Horv93}
\begin{align}\label{ShrAlN1}
&  \{Q,Q\}=2iH, && \{Q,\Lambda_i\}=iP_i && [H,S]=Q,\nn
\\
& \{Q,S\}=2iD, && \{S,\Lambda_i\}=iC_i, && [D,S]=\frac12S,\nn
\\
& \{S,S\}=2iK, && \{\Lambda_i,\Lambda_j\}=i\delta_{ij}m, &&
[D,Q]=-\frac12Q,\nn
\\
& [K,Q]=-S, && [Q,C_i]=\Lambda_i, && [S,P_i]=-\Lambda_i,\nn
\\
&
[\Lambda_i,M_{jk}^{}]=\delta_{ij}\Lambda_{k}-\delta_{ik}\Lambda_{j},
&&
\end{align}
As a sub-superalgebra it contains $osp(1|1)$ spanned by $H,D,K$ and
$Q,S$.

An $N=2$ Schr\"odinger superalgebra includes
pairs of Grassmann-odd generators describing the supercharges  $Q,\bar{Q}$,
the generators of
superconformal transformations $S,\bar S$, the superpartners
$\Lambda_i,\bar\Lambda$ of the Galilei boosts, as well as extra
Grassmann-even generator $J$, which corresponds to $u(1)$
$R$-symmetry. Additional (anti)-commutation relations read
\cite{GM09}
\begin{align}\label{ShrAlN2}
&  \{Q,\bar Q\}=2iH, && \{\bar Q,\Lambda_i\}=iP_i, && [H,S]=Q,\nn
\\
& \{Q,\bar S\}=2iD+J+Z, && \{S,\bar\Lambda_i\}=iC_i, &&
[D,S]=\frac12S,\nn
\\
& \{S,\bar S\}=2iK, && \{\Lambda_i,\bar\Lambda_j\}=i\delta_{ij}m, &&
[D,Q]=-\frac12Q,\nn
\\
& [K,Q]=-S, && [Q,C_i]=\Lambda_i, && [S,P_i]=-\Lambda_i,\nn
\\
& [K,\bar Q]=-\bar S,  && \{ Q,\bar\Lambda_i\}=iP_i, && [H,\bar
S]=\bar Q,\nn
\\
& \{\bar Q, S\}=2iD-J-Z, && \{\bar S,\Lambda_i\}=iC_i, && [D,\bar
S]=\frac12\bar S,\nn
\\
& [\bar Q,C_i]=\bar\Lambda_i, && [\bar S,P_i]=-\bar\Lambda_i, &&
[D,\bar Q]=-\frac12\bar Q,\nn
\\
& [J,Q]=iQ, && [J,\bar Q]=-i\bar Q,  && [J,S]=iS,\nn
\\
& [J,\bar S]=-i\bar S, && [J,\Lambda_i]=i\Lambda_i, &&
[J,\bar\Lambda_i]=-i\bar\Lambda_i,\nn
\\
&
[\Lambda_i,M_{jk}^{}]=\delta_{ij}\Lambda_{k}-\delta_{ik}\Lambda_{j},
&&
[\bar\Lambda_i,M_{jk}^{}]=\delta_{ij}\bar\Lambda_{k}-\delta_{ik}\bar\Lambda_{j},
&&
\end{align}
where $Z$ is a central charge. Generators $H,D,K,J$ and
$Q,\bar{Q},S,\bar{S}$ form $usp(1,1|1)$ superalgebra. The $N=2$
Schr\"odinger supersymmetry was revealed when studying the model of
a harmonic oscillator extended by fermionic degrees of freedom
\cite{BH86}. In compact notation, the structure relations
(\ref{ShrAlN1}) and (\ref{ShrAlN2}) are given in Appendix. For
further discussion on the $N=1,2$ Schr\"odinger superalgebras see
\cite{Horv93,GM09,DH94,HU05}.

\section{Perfect fluid dynamics}\label{Sec3}

In non-relativistic space-time parameterized by the coordinates
$(t,x_i)$, $i=1,...,d$, a perfect fluid is characterized by the
density  $\rho(t,x)$ and the velocity vector field
$\upsilon_i(t,x)$. Its evolution over time is described by the
continuity equation for density and the Euler
equation\footnote{Throughout the text, we use the notations:
$\partial_0=\frac{\partial}{\partial t}$,
$\partial_i=\frac{\partial}{\partial x_i}$, ${\cal
D}=\partial_0+\upsilon_i\partial_i$. Summation over repeated indices
is understood. Considering the coordinates $t$ and $x_i$ as
independent, one has the identity ${\cal D}x_i=\upsilon_i$.}
\begin{eqnarray}\label{PFEq}
{\partial_0\rho}+ {\partial_i (\rho\upsilon_i)}=0,\quad {\cal
D}\upsilon_i=-\frac{1}{\rho}{\partial_i p},
\end{eqnarray}
where ${\cal D}=\partial_0+\upsilon_i\partial_i$ is the material
derivative and $p(t,x)$ is the pressure which is assumed to be
related to the density via an equation of state $p=p(\rho)$.

The equations of motion (\ref{PFEq}) can be obtained from the Hamiltonian
\begin{eqnarray}\label{Ham}
H=\int \rmd
x\left(\frac12\rho\upsilon^{}_i\upsilon^{}_i+V\right),\quad \rmd
x=\rmd x_1\rmd x_2...\rmd x_d,
\end{eqnarray}
provided the non-canonical Poisson brackets were introduced
\cite{MG80}
\begin{eqnarray}\label{PBPF}
\{\rho(x),\upsilon_i(x')\}&=&- {\partial_i}\delta(x,x'), \nn\\
\{\upsilon_i(x),\upsilon_j(x')\}&=&\frac{1}{\rho}\omega_{ij}\delta(x,x')=\frac{1}{\rho}\left({\partial_i\upsilon_j}
-{\partial_j\upsilon_i}\right)\delta(x,x'),
\end{eqnarray}
where $\omega_{ij}$ is called the fluid vorticity. The potential
function $V(\rho)$ in (\ref{Ham}) links to the pressure via the
Legendre transformation $p=\rho V'-V$ (prime denotes differentiation
with respect to the argument). $V'$ is called the enthalpy and
$c=\sqrt{p'}=\sqrt{\rho V''}$ is the speed of sound which depends on
the density $\rho(t,x)$.

The non-relativistic perfect fluid equations can alternatively be
formulated as the conservation of the energy-momentum tensor (see e.g.
\cite{JNPP04})
\begin{align}
&T^{00}=\frac12\rho\upsilon_i\upsilon_i+V, &&
T^{i0}=\rho\upsilon_i(\frac12\upsilon_j\upsilon_j+V')\nn
\\
&T^{0i}=\rho\upsilon_i, &&
T^{ji}=\rho\upsilon_i\upsilon_j+\delta_{ij}p,
\end{align}
where the components $T^{00}$ and $T^{i0}$ are identified with the
energy density and the energy flux density whereas $T^{i0}$ and
$T^{ji}$ link to the momentum density and the stress tensor. They
satisfy the equations
\begin{eqnarray}\label{Tdifeq}
{\partial_0 T^{00}}+{\partial_i T^{i0}}=0,\qquad {\partial_0
T^{0i}}+{\partial_j T^{ji}}=0,
\end{eqnarray}
as well as the algebraic condition
\begin{eqnarray}\label{Talgeq}
2T^{00}=\delta_{ij}T^{ij},\qquad V=\frac12dp,
\end{eqnarray}
where $dp$ is a product of the spatial dimension $d$ and the
pressure $p$. Two comments are in order. Firstly, $T^{i0}\neq
T^{0i}$ because the theory is not Lorentz-invariant but $T^{ij}=
T^{ji}$ because it is invariant under spatial rotations. Secondly,
the condition (\ref{Talgeq}) is the analogue of the tracelessness
condition characterizing a relativistic conformal field theory. It
is satisfied only for $p\sim\rho^{1+\frac{2}{d}}$, where $d$ is the
spatial dimension.

The equations (\ref{Tdifeq}), the condition (\ref{Talgeq}) and the
properties of the energy-momentum tensor allow one to construct
integrals of motion that correspond to symmetries of the theory (for
a related discussion see \cite{HU03})
\begin{eqnarray}
H&=&\int \rmd x T^{00}=\int \rmd
x(\frac12\rho\upsilon_i\upsilon_i+V),\label{IM1}
\\
P_i^{}&=&\int \rmd x T^{0i}=\int \rmd x \rho
\upsilon^{}_i,\label{IM2}
\\
C_i^{}&=&\int \rmd x (T^{0i}t-\rho x_i)=tP_i-\int \rmd x \rho
x_i,\label{IM3}
\\
M_{ij}&=&\int \rmd x (T^{0i}x_j-T^{0j}x_i)=\int \rmd x
(\rho\upsilon_ix_j-\rho\upsilon_jx_i),\label{IM4}
\\
D&=&\int \rmd x (T^{00}t-\frac12 T^{0i}x_i)=tH-\frac12\int \rmd
x\rho\upsilon_i^{}x_i,\label{IM5}
\\
K&=&\int \rmd x (T^{00}t^2- T^{0i}tx_i+\frac12\rho x_ix_i)=
-t^2H+2tD+\frac12\int \rmd x\rho x_ix_i.\label{IM6}
\end{eqnarray}
where $H$, $P_i$, $M_{ij}$, $C_i$, $D$, and $K$ are associated with
the temporal translation, spatial translation, spatial rotations,
Galilei boosts, dilation and special conformal transformation,
respectively. In order to verify the conservation of (\ref{IM3}) and
(\ref{IM6}) over time, one should also use the continuity equation
for the density $\partial_0\rho+\partial_iT^{0i}=0$. Under the
Poisson bracket, the integrals of motion (\ref{IM1})-(\ref{IM6})
satisfy the structure relations of the Schr\"odinger algebra
(\ref{ShrAl}) and generate the corresponding symmetry
transformations for $\rho$ and $\upsilon_i$. The role of the central
charge is played by the total mass $m=\int \rmd x\rho$, which is
conserved due to the continuity equation for the density.

\section{Perfect fluid dynamics with $N=1$ Schr\"odinger supersymmetry}\label{Sec4}

When constructing supersymmetric extensions below,
we assume the fluid pressure to obey the polytropic
equation of state
\begin{eqnarray}\label{pr}
p=\lambda(\gamma-1)\rho^\gamma,\quad \gamma\neq0,1,
\end{eqnarray}
where $\lambda>0$ is a constant and $\gamma$ is the polytropic
exponent. A
large class of physical systems is characterized
by such an equation of state. In particular, at $\gamma=-1$ one reveals
the Chaplygin
gas \cite{chap1904}. As was explained above, for
$\gamma=1+\frac{2}{d}$, where $d$ is the spatial dimension, the
perfect fluid equations (\ref{PFEq}) hold invariant under the
Schr\"odinger group.

For what follows, it is worth mentioning that, given the equation of state (\ref{pr}),
the Legendre transformation $p=\rho V'-V$ yields
the potential
\begin{eqnarray}\label{PolPot}
V=\frac{1}{\gamma-1}p=\lambda\rho^\gamma,
\end{eqnarray}
which enters the Hamiltonian (\ref{Ham}). The latter
leads to the speed of sound $c=\sqrt{\rho
V''}=\sqrt{\lambda\gamma(\gamma-1)\rho^{\gamma-1}}$, which reduces to
$$c=\frac{\sqrt{2\lambda(d+2)}}{d}\rho^{\frac{1}{d}}$$ in the Schr\"odinger invariant case ($\gamma=1+\frac{2}{d}$).

In order to construct an $N=1$ supersymmetric extension, let us
introduce real Grassmann-odd partners\footnote{We use the
conventional notation for complex conjugation of Grassmann-odd
variables $\psi_1$ and $\psi_2$:
$(\psi_1\psi_2)^*=\psi_2^*\psi_1^*$. In case of real fermions
$\psi^*=\psi$, one obtains $(\psi_1\psi_2)^*=-\psi_1\psi_2$. }
$\sigma(t,x)$ and $\psi_i(t,x), i=1,...,d$ for the Grassmann-even
fields characterizing the density $\rho(t,x)$ and the velocity
$\upsilon_i(t,x)$, respectively. We also impose the following
Poisson brackets:
\begin{eqnarray}\label{PBN1_1}
\{\psi_i(x),\psi_j(x')\}=\frac{i}{\rho}\delta_{ij}\delta(x,x'),\qquad
\{\sigma(x),\sigma(x')\}=\frac{i}{\rho}\delta(x,x').
\end{eqnarray}
In order to ensure the super-Jacobi identities, one has to require
\begin{eqnarray}\label{PBN1_2}
\{\upsilon_i(x),\psi_j(x')\}=\frac{1}{\rho}{\partial_i\psi_j}\delta(x,x'),\qquad
\{\upsilon_i(x),\sigma(x')\}=\frac{1}{\rho}\partial_i\sigma\delta(x,x').
\end{eqnarray}
Within the context of supersymmetric fluid
mechanics, similar Poisson brackets have previously been proposed in \cite{JP00}.

Given (\ref{PBN1_1}) and (\ref{PBN1_2}), an $N=1$ supersymmetry
charge can be chosen in the simplest form, which is linear in the
Grassmann-odd variables
\begin{eqnarray}
Q=\int \rmd x
(\rho\upsilon_i\psi_i+\sqrt{2\lambda}\rho^{\frac{\gamma+1}{2}}\sigma).
\end{eqnarray}
Via the Poisson bracket, it gives rise to the super-extended
Hamiltonian
\begin{eqnarray}\label{HamN1}
\{Q,Q\}=i\int \rmd x
\Big(\rho\upsilon_i\upsilon_i+2\lambda\rho^{\gamma}-i\sqrt{2\lambda}(\gamma-1)\rho^\frac{\gamma+1}{2}\partial_i\psi_i\sigma\Big)=2iH.
\end{eqnarray}
The first two terms entering the last expression reproduce
the original bosonic theory (\ref{Ham}), while the last contribution describes
the boson-fermion coupling. From the super-Jacobi identities it follows
that the supercharge is conserved,
$\{Q,H\}=0$.

It is worth mentioning that, although the potential (\ref{PolPot}) was
explicitly used above, the supersymmetric construction works for an
arbitrary $V(\rho)$. In general, the supercharge can be chosen in the form
\begin{eqnarray}
Q=\int \rmd x (\rho\upsilon_i\psi_i+\sqrt{2\rho V}\sigma),
\end{eqnarray}
and the super-extended Hamiltonian then then reads
form
\begin{eqnarray}
H=\int \rmd x
\Big(\frac12\rho\upsilon_i\upsilon_i+V-i\frac{\rho(\rho
V'-V)}{\sqrt{2\rho V}}\partial_i\psi_i\sigma\Big),
\end{eqnarray}
which for $V=\lambda\rho^{\gamma}$ reproduces
(\ref{HamN1}).

The super-extended Hamiltonian (\ref{HamN1}) generates the $N=1$
supersymmetric dynamics characterized by the following equations of motion
\begin{eqnarray}
\partial_0\rho&=&\{\rho,H\}=-\partial_i(\rho\upsilon_i),\label{N1eq1}
\\
\partial_0\sigma&=&\{\sigma,H\}=-\upsilon_i\partial_i\sigma-\frac{\sqrt{2\lambda}}{2}(\gamma-1)\rho^{\frac{\gamma-1}{2}}\partial_i\psi_i,\label{N1eq2}
\\
\partial_0\psi_i&=&\{\psi_i,H\}=-\upsilon_j\partial_j\psi_i-\frac{\sqrt{2\lambda}}{2}(\gamma-1)\frac{1}{\rho}\partial_i(\rho^{\frac{\gamma+1}{2}}\sigma),\label{N1eq3}
\\
\partial_0\upsilon_i&=&\{\upsilon_i,H\}=-\upsilon_j\partial_j\upsilon_i-\lambda(\gamma-1)\frac{1}{\rho}\partial_i(\rho^{\gamma})\nn\\
&&+i\frac{\sqrt{2\lambda}}{4}(\gamma-1)\frac{1}{\rho}\Big((\gamma-1)\partial_i(\rho^{\frac{\gamma+1}{2}}\partial_j\psi_j\sigma)
+2\partial_j(\rho^{\frac{\gamma+1}{2}}\partial_i\psi_j\sigma)\Big).\label{N1eq4}
\end{eqnarray}

In order to analyze symmetries of the model under consideration, it proves instructive
 to write down components of the super-extended
energy-momentum tensor
\begin{eqnarray*}
T^{00}&=&\frac12\rho\upsilon_i\upsilon_i+\lambda\rho^{\gamma}-i\frac{\sqrt{2\lambda}}{2}(\gamma-1)\rho^\frac{\gamma+1}{2}\partial_i\psi_i\sigma,
\\
T^{i0}&=&\rho\upsilon_i\Big(\frac12\upsilon_j\upsilon_j+\lambda\gamma\rho^{\gamma-1}
-i\frac{\sqrt{2\lambda}}{4}(\gamma+1)(\gamma-1)\rho^{\frac{\gamma-1}{2}}\partial_j\psi_j\sigma\Big)\\
&&-i\frac{\sqrt{2\lambda}}{2}(\gamma-1)\rho^{\frac{\gamma+1}{2}}\upsilon_j\partial_j\psi_i\sigma+i\frac{\lambda}{2}(\gamma-1)^2\rho^{\gamma}\sigma\partial_i\sigma,
\\
T^{0i}&=&\rho\upsilon_i,
\\
T^{ji}&=&\rho\upsilon_i\upsilon_j+\delta_{ij}\lambda(\gamma-1)\rho^{\gamma}
-i\frac{\sqrt{2\lambda}}{4}(\gamma-1)\Big((\gamma-1)\delta_{ij}\rho^\frac{\gamma+1}{2}\partial_k\psi_k\sigma+2\rho^\frac{\gamma+1}{2}\partial_i\psi_j\sigma\Big).
\end{eqnarray*}
If the equations of motion (\ref{N1eq1})-(\ref{N1eq4}) hold, the energy-momentum tensor obeys
$\partial_\mu T^{\mu\nu}=0$, $\mu=(0,i)$,
which ensures the conservation of the total energy, total momentum and
charges corresponding to the Galilei boosts
\begin{eqnarray}
H=\int \rmd x T^{00},\quad P_i=\int \rmd x T^{0i},\quad C_i=\int
\rmd x (T^{0i}t-\rho x_i).
\end{eqnarray}
In the special case $\gamma=1+\frac{2}{d}$, the
tracelessness-like condition $2T^{00}=\delta_{ij}T^{ij}$ is
satisfied, which allows one to construct conserved charges associated with the
dilation and special conformal transformation
\begin{eqnarray}
D=\int \rmd x (T^{00}t-\frac12 T^{0i}x_i),\quad  K=\int \rmd x
(T^{00}t^2- T^{0i}tx_i+\frac12\rho x_ix_i).
\end{eqnarray}
As far as the angular momentum is concerned, one cannot directly use
the formula (\ref{IM4}) because the stress tensor $T^{ji}$ is not
symmetric. However, it can be symmetrized by taking into account the form of $T^{0i}$ and
adding a
total derivative term. The improved quantities read
\begin{eqnarray}
{T}_I^{0i}&=&\rho\upsilon_i+\frac{i}{2}\partial_k(\rho\psi_i\psi_k),\label{TN1Im1}
\\
{T}_I^{ji}&=&\rho\upsilon_i\upsilon_j+\delta_{ij}\lambda(\gamma-1)\rho^{\gamma}\nn\\
&&
-i\frac{\sqrt{2\lambda}}{4}(\gamma-1)\Big((\gamma-1)\delta_{ij}\rho^\frac{\gamma+1}{2}\partial_k\psi_k\sigma
+\rho^\frac{\gamma+1}{2}\partial_i\psi_j\sigma+\rho^\frac{\gamma+1}{2}\partial_j\psi_i\sigma\Big)\nn\\
&&+\frac{i}{2}\partial_k(\rho\upsilon_i\psi_j\psi_k+\rho\upsilon_j\psi_i\psi_k)\nn\\
&&+i\frac{\sqrt{2\lambda}}{4}(\gamma-1)\Big(\partial_i(\rho^{\frac{\gamma+1}{2}}\psi_j\sigma)+
\partial_j(\rho^{\frac{\gamma+1}{2}}\psi_i\sigma)-2\delta_{ij}\partial_k(\rho^{\frac{\gamma+1}{2}}\psi_k\sigma)\Big),\label{TN1Im2}
\end{eqnarray}
which obey the desired equation ${\partial_0 T_I^{0i}}+{\partial_j T_I^{ji}}=0$. Then it
immediately follows that the angular momentum
\begin{eqnarray}
M_{ij}&=&\int \rmd x (T_I^{0i}x_j-T_I^{0j}x_i)=\int \rmd x
(\rho\upsilon_ix_j-\rho\upsilon_jx_i)-i\int \rmd x \rho\psi_i\psi_j
\end{eqnarray}
is conserved. The first term is the orbital angular momentum while
the second is the spin part. Note that the extra term in the
improved momentum density (\ref{TN1Im1})
$\frac{i}{2}\partial_k(\rho\psi_i\psi_k)$ leaves the total momentum
$P_i=\int \rmd x {T}_I^{0i}$ unchanged after integration (surface
terms are ignored). In the literature, it is interpreted as an
additional localized momentum density responsible for the spin part
of the angular momentum.

Let us turn to supersymmetries. In addition to the
previously constructed supercharge $Q$, it is easy to verify that the Grassmann-odd
vector quantity
\begin{eqnarray}
\Lambda_i=\int \rmd x \rho\psi_i
\end{eqnarray}
is conserved.

It proves useful to relate the conservation of $Q$ and $\Lambda_i$
with the corresponding continuity equations. To do this, we
introduce an Grassmann-odd tensor $Y^{\mu,\nu}$, $\mu=(0,i)$ with the
components
\begin{eqnarray}
Y^{0,0}&=&\rho\upsilon_i\psi_i+\sqrt{2\lambda}\rho^{\frac{\gamma+1}{2}}\sigma,\nn
\\
Y^{i,0}&=&\rho\upsilon_i\Big(\upsilon_j\psi_j+\frac{\sqrt{2\lambda}}{2}(\gamma+1)\rho^{\frac{\gamma-1}{2}}\sigma\Big)+
\lambda(\gamma-1)\rho^{\gamma}\psi_i\nn\\
&&
+i\frac{\sqrt{2\lambda}}{4}(\gamma-1)\Big((\gamma-1)\rho^{\frac{\gamma+1}{2}}\partial_j\psi_j\psi_i\sigma+2\rho^{\frac{\gamma+1}{2}}\partial_j\psi_i\psi_j\sigma\Big),\nn
\\
Y^{0,i}&=&\rho\psi_i,\nn
\\
Y^{j,i}&=&\rho\upsilon_j\psi_i+\frac{\sqrt{2\lambda}}{2}(\gamma-1)\delta_{ij}\rho^{\frac{\gamma+1}{2}}\sigma,
\end{eqnarray}
which obey
\begin{eqnarray}
{\partial_0 Y^{0,0}}+{\partial_i Y^{i,0}}=0,\qquad {\partial_0
Y^{0,i}}+{\partial_j Y^{j,i}}=0.
\end{eqnarray}
The tensor $Y^{\mu,\nu}$ plays the role of a superpartner for the
energy-momentum tensor $T^{\mu\nu}$ and its components are
interpreted as the densities and fluxes for $Q$ and $\Lambda_i$ via
the identification
\begin{eqnarray}
Q=\int \rmd x Y^{0,0},\qquad \Lambda_i=\int \rmd x Y^{0,i}.
\end{eqnarray}

Note that, similarly to the energy-momentum tensor, one reveals the
tracelessness-like condition for the same value of the polytropic
exponent
\begin{eqnarray}
Y^{0,0}=\delta_{ij}Y^{i,j},\qquad \gamma=1+\frac{2}{d}.
\end{eqnarray}
This gives an additional conserved quantity
\begin{eqnarray}
S=\int \rmd x (Y^{0,0}t- Y^{0,i}x_i)=Qt-\int \rmd x\rho\psi_ix_i
\end{eqnarray}
associated with the superconformal transformation generator. Under the Poisson bracket,
the
Grassmann-odd conserved charges $Q$, $\Lambda_i$, and $S$ obey the
structure relations of the $N=1$ Schr\"odinger
superalgebra (\ref{ShrAlN1}).

Thus, we have constructed a non-relativistic perfect fluid system
with the $N=1$ Schr\"odinger supersymmetry which is governed by the
super-extended Hamiltonian (\ref{HamN1}) and the non-canonical
Poisson brackets (\ref{PBPF}),(\ref{PBN1_1}), and (\ref{PBN1_2}).
The corresponding field variables include the Grassmann-even density
and velocity $(\rho,\upsilon_i)$ and their Grassmann-odd
super-partners $(\sigma,\psi_i)$. In the bosonic limit
$(\sigma,\psi_i)\rightarrow0$ the original non-supersymmetric theory
is reproduced. As a new result, we provide a complete set of
conserved charges forming the $N=1$ Schr\"odinger superalgebra, all
of which can be derived from the the super-extended stress-energy
tensor $T^{\mu\nu}$ and its Grassmann-odd superpartner
$Y^{\mu,\nu}$.

Note that similar
anticommuting field variables arise in the component decomposition
of the superfield hydrodynamics with the $N=1$ Schr\"odinger
supersymmetry in \cite{Gal24}. However, in our our approach there are
no extra bosonic fields and fewer fermionic fields.

Concluding this section, we note that assigning a physical meaning to anticommuting
variables in fluid mechanics is usually problematic. Nevertheless,
for the case under consideration, one
can consistently interpret  $(\sigma,\psi_i)$ if one switches to the
Lagrangian formulation determined by
\begin{eqnarray}\label{LagN1}
L&=&-\int \rmd
x\rho(\partial_0\theta+\frac{i}{2}\psi_j\partial_0\psi_j+\frac{i}{2}\sigma\partial_0\sigma)-H\nn
\\
&=&-\int
\rmd x\rho(\partial_0\theta+\frac{i}{2}\psi_j\partial_0\psi_j+\frac{i}{2}\sigma\partial_0\sigma)\nn\\
&&\qquad\qquad-\int \rmd x
\Big(\frac12\rho\upsilon_i\upsilon_i+\lambda\rho^{\gamma}-\frac{i}{2}\sqrt{2\lambda}(\gamma-1)\rho^\frac{\gamma+1}{2}\partial_i\psi_i\sigma\Big),
\end{eqnarray}
where $H$ is the super-extended Hamiltonian (\ref{HamN1}) and the velocity vector field
$\upsilon_i$ is represented as the Clebsch-like
decomposition
\begin{eqnarray}\label{Cleb}
\upsilon_i=\partial_i\theta+\frac{i}{2}\psi_j\partial_i\psi_j+\frac{i}{2}\sigma\partial_i\sigma
\end{eqnarray}
with Grassmann-even prepotential $\theta$. Supersymmetric
equations of motion then follow from the variational principle. The variation
with respect to $\theta$ gives the continuity equation for the
density (\ref{N1eq1}). The variations with
respect to $\sigma$ and $\psi$  give the equations (\ref{N1eq2}) and
(\ref{N1eq3}), and from the variation with respect to $\rho$ one
obtains
\begin{eqnarray}
{\cal
D}\theta=\frac12\upsilon_i\upsilon_i-\lambda\gamma\rho^{\gamma-1}+\frac{i}{4}\sqrt{2\lambda}(\gamma-1)\Big(\gamma\rho^{\frac{\gamma-1}{2}}\partial_i\psi_i\sigma
+\frac{1}{\rho}\psi_i\partial_i(\rho^{\frac{\gamma+1}{2}}\sigma)\Big),
\end{eqnarray}
where (\ref{N1eq1})-(\ref{N1eq3}) were used  and ${\cal
D}=\partial_0+\upsilon_i\partial_i$ is the material derivative. As a
result, the super-extended Euler equation (\ref{N1eq4}) is
reproduced
\begin{eqnarray}
{\cal D}\upsilon_i&=&{\cal
D}(\partial_i\theta+\frac{i}{2}\psi_j\partial_i\psi_j+\frac{i}{2}\sigma\partial_i\sigma)\nn
\\
&=&-\lambda(\gamma-1)\frac{1}{\rho}\partial_i(\rho^{\gamma})\nn\\
&&
+i\frac{\sqrt{2\lambda}}{4}(\gamma-1)\frac{1}{\rho}\Big((\gamma-1)\partial_i(\rho^{\frac{\gamma+1}{2}}\partial_j\psi_j\sigma)
+2\partial_j(\rho^{\frac{\gamma+1}{2}}\partial_i\psi_j\sigma)\Big).
\end{eqnarray}

Thus, (\ref{Cleb}) is a Clebsch-like decomposition of the velocity
vector field in which the anticommuting variables $(\sigma,\psi_i)$
play the role of the "Gaussian" potentials giving rise to the
non-vanishing fluid's vorticity
\begin{eqnarray}
\omega_{ij}=\partial_{i}\upsilon_{j}-\partial_{j}\upsilon_{i}=i(\partial_{i}\psi_k\partial_j\psi_k+\partial_{i}\sigma\partial_j\sigma).
\end{eqnarray}
In the bosonic limit $(\sigma,\psi_i)\rightarrow0$ the vorticity
vanishes and the Lagrangian (\ref{LagN1}) describes the dynamics of
an irrotational fluid $\upsilon_i=\partial_i\theta$. The possibility
to parameterize fluid's vorticity by anticommuting variables was
first proposed in \cite{JP00}, where the supersymmetric planar
Chaplygin gas model was constructed.

\section{Perfect fluid dynamics with $N=2$ Schr\"odinger supersymmetry}\label{Sec5}

In order to construct an $N=2$ supersymmetric extension of the
perfect fluid equations, we introduce complex anticommuting fields
$(\sigma,\psi_i)$ as superpartners for the density and velocity
$(\rho,\upsilon_i)$ and impose the following Poisson brackets
\begin{align}\label{PBN2}
&\{\psi_i(x),\bar\psi_j(x')\}=\frac{i}{\rho}\delta_{ij}\delta(x,x'),
&&\{\upsilon_i(x),\psi_j(x')\}=\frac{1}{\rho}{\partial_i\psi_j}\delta(x,x'),\nn
\\
& \{\sigma(x),\bar\sigma(x')\}=\frac{i}{\rho}\delta(x,x'), &&
\{\upsilon_i(x),\sigma(x')\}=\frac{1}{\rho}\partial_i\sigma\delta(x,x'),\nn
\\
&\{\psi_i(x),\psi_j(x')\}=0,
&&\{\upsilon_i(x),\bar\psi_j(x')\}=\frac{1}{\rho}{\partial_i\bar\psi_j}\delta(x,x'),\nn
\\
& \{\sigma(x),\sigma(x')\}=0, &&
\{\upsilon_i(x),\bar\sigma(x')\}=\frac{1}{\rho}\partial_i\bar\sigma\delta(x,x'),
\end{align}
where $(\bar\sigma,\bar\psi_i)$ are complex conjugates of
$(\sigma,\psi_i)$. It is straightforward to verify that the super-Jacobi identities
are satisfied.

As the nest step, two self-adjoint supercharge generators are constructed
\begin{eqnarray}\label{SupN2}
Q&=&\int \rmd x(
\rho\upsilon_i\psi_i+\sqrt{2\lambda}\rho^{\frac{\gamma+1}{2}}\sigma+\frac{i}{2}(\gamma-1)\rho\partial_i\psi_i\sigma\bar\sigma),\nn
\\
\bar Q&=&\int \rmd x
(\rho\upsilon_i\bar\psi_i+\sqrt{2\lambda}\rho^{\frac{\gamma+1}{2}}\bar\sigma-\frac{i}{2}(\gamma-1)\rho\partial_i\bar\psi_i\sigma\bar\sigma),
\end{eqnarray}
which via $\{Q,\bar Q\}=2iH$ give rise to the $N=2$ super-extended Hamiltonian
\begin{eqnarray}\label{HamN2}
H&=&\int \rmd x
\Big(\frac12\rho\upsilon_i\upsilon_i+\lambda\rho^{\gamma}-i\frac{\sqrt{2\lambda}}{2}(\gamma-1)\rho^\frac{\gamma+1}{2}(\partial_i\psi_i\bar\sigma+\partial_i\bar\psi_i\sigma)\nn\\
&&\qquad\qquad
-\frac12(\gamma-1)\rho\partial_i\psi_j\partial_j\bar\psi_i\sigma\bar\sigma
+\frac{1}{8}(\gamma-1)^2\frac{1}{\rho}\partial_i(\rho\sigma\bar\sigma)\partial_i(\rho\sigma\bar\sigma)\Big).
\end{eqnarray}
In contrast to the $N=1$ case studied above, the supercharges involve terms cubic in
the fermionic fields which are required by $\{Q,Q\}=0$
and $\{\bar Q,\bar Q\}=0$.

From the super-extended Hamiltonian one obtains the equations of motion
\begin{eqnarray}
{\cal D}\rho&=&-\rho\partial_i\upsilon_i,
\\
{\cal
D}\sigma&=&-\frac{\sqrt{2\lambda}}{2}(\gamma-1)\rho^{\frac{\gamma-1}{2}}\partial_i\psi_i
+i\frac{\gamma-1}{2}\partial_i\psi_j\partial_j\bar\psi_i\sigma
+\frac{i}{4}(\gamma-1)^2\sigma\partial_i(\frac{1}{\rho}\partial_i(\rho\sigma\bar\sigma)),\label{N2eq_2}
\\
\rho{\cal
D}\psi_i&=&-\frac{\sqrt{2\lambda}}{2}(\gamma-1)\partial_i(\rho^{\frac{\gamma+1}{2}}\sigma)
-i\frac{(\gamma-1)}{2}\partial_j(\rho\partial_i\psi_j\sigma\bar\sigma),\label{N2eq_3}
\\
\rho{\cal
D}\upsilon_i&=&-(\gamma-1)\partial_i\Big(\lambda\rho^{\gamma}
-i\frac{\sqrt{2\lambda}}{4}(\gamma-1)\rho^{\frac{\gamma+1}{2}}(\partial_j\bar\psi_j\sigma
+\partial_j\psi_j\bar\sigma)\Big)\nn\\
&&
+\frac12(\gamma-1)\partial_j\Big(i\sqrt{2\lambda}\rho^{\frac{\gamma+1}{2}}(\partial_i\bar\psi_j\sigma+\partial_i\psi_j\bar\sigma)
+\rho\partial_i\psi_k\partial_k\bar\psi_j\sigma\bar\sigma\nn\\
&&\qquad\qquad\qquad\qquad -
\rho\partial_i\bar\psi_k\partial_k\psi_j\sigma\bar\sigma
+\frac{(\gamma-1)}{2}\rho\sigma\bar\sigma\partial_i(\frac{1}{\rho}\partial_j(\rho\sigma\bar\sigma))\Big),
\end{eqnarray}
the first of which is the continuity equation for the density rewritten
in terms of the material derivative ${\cal
D}=\partial_o+\upsilon_i\partial_i$. The equations of motion for
$\bar\sigma$ and $\bar\psi_i$ are obtained from (\ref{N2eq_2}) and
(\ref{N2eq_3}) by the complex conjugation.

The integrals of motion realizing the Grassmann-even part of the
$N=2$ Schr\"odinger superalgebra reads
\begin{eqnarray}\label{N2IM_ev}
P_i^{}&=&\int \rmd x \rho \upsilon^{}_i,\qquad C_i^{}=tP_i-\int \rmd
x \rho x_i,\qquad J=\int \rmd x
\rho(\psi_i\bar\psi_i+\sigma\bar\sigma)\nn
\\
M_{ij}&=&\int \rmd x (\rho\upsilon_ix_j-\rho\upsilon_jx_i)-i\int
\rmd x \rho(\psi_i\bar\psi_j-\psi_j\bar\psi_i),\nn
\\
D&=&tH-\frac12\int \rmd x\rho\upsilon_i^{}x_i,\qquad K=
-t^2H+2tD+\frac12\int \rmd x\rho x_ix_i,
\end{eqnarray}
where $H$ is time-independent Hamiltonian (\ref{HamN2}). The last
two integrals of motion are present for a particular value
of the polytropic exponent $\gamma=1+\frac{2}{d}$ only. Note that,
in contrast to the $N=1$ case, there is an additional Grassmann-even
integral of motion $J$ associated with the $u(1)$ $R$-symmetry.

The Grassmann-odd part of the $N=2$ Schr\"odinger superalgebra is
realized by the supercharges $Q$ and $\bar Q$, as well as by the
integrals of motion
\begin{align}\label{N2IM_odd}
& \Lambda_i=\int \rmd x \rho\psi_i, && \bar\Lambda_i=\int \rmd x
\rho\bar\psi_i,\nn
\\
& S=Qt-\int \rmd x\rho\psi_ix_i, && \bar S=\bar Qt-\int \rmd
x\rho\bar\psi_ix_i,
\end{align}
where $S,\bar{S}$ correspond the superconformal generators and
$\Lambda_i,\bar\Lambda_i$ corresponds to the superpartners of the Galilei
boosts.

It is straightforward to verify that the conserved charges
(\ref{SupN2}),(\ref{HamN2}), (\ref{N2IM_ev}), and (\ref{N2IM_odd})
 obey the structure relations of the
$N=2$ Schr\"odinger superalgebra (\ref{ShrAlN2}) with central
charges $m=\int \rmd x\rho$ and $Z=0$.

\section{Conclusion}\label{Sec6}

To summarize, in this work the perfect
fluid equations with the $N=1,2$
Schr\"odinger supersymmetry were formulated within the Hamiltonian
formalism. This was achieved by introducing the Grassmann-odd superpartners for the
density and velocity and building the corresponding
supersymmetry charges in the extended field space. The supersymmetry charges
generate the super-extended Hamiltonian via the Poisson bracket. The super-extended Hamiltonian
governs the dynamics of the resulting fluid. The supercharge is linear in the fermionic
fields for $N=1$ and it is cubic for $N=2$. The full set of conserved
charges associated with the $N=1,2$ Schr\"odinger
superalgebra was built. For the $N=1$ case, it was demonstrated that
the anticommuting variables can be
interpreted as potentials parameterizing fluid's vorticity.
The perfect
fluid equations with the $N=2$
Schr\"odinger supersymmetry
constructed in this work are essentially new.

The present work was mostly focused on the mathematical structure of
fluid mechanics with the $N=1,2$ Schr\"odinger supersymmetries.
Turning to possible physical applications, it is tempting to suggest
that the Grassmann-odd partners might represent spin degrees of
freedom. In this regard a possible link to micropolar fluid
mechanics \cite{Eri1966}, where each fluid element carries intrinsic
spin, is interesting to explore. As was demonstrated above, in the
$N=2$ case the anticommuting variables can be interpreted as
potentials parameterizing fluid's vorticity. The related physics
deserves a separate investigation by constructing explicit solutions
to the equations of motion using the procedure in \cite{GH11,GH13}
and analyzing their qualitative dynamics. Also it would be
interesting to understand whether the supersymmetric fluid equations
(\ref{N1eq1})-(\ref{N1eq4}) with the $N=1$ Schr\"odinger
supersymmetry can be related to the superfield formulation in
\cite{Gal24}. The construction of $N=3,4$ supersymmetric extensions
is an intriguing open problem. The generalizations of the present
analysis to fluid models with the $\ell$-conformal Galilei symmetry
\cite{Gal22a,Gal22b,Sne23a,Sne24}, the Lifshitz symmetry
\cite{Gal22b}, or the conformal Newton-Hooke symmetry \cite{Sne25a}
is an interesting avenue to explore.

\section*{Acknowledgements}
The author thanks an anonymous reviewer for comments and questions
which helped to improve the manuscript. This work was supported by
the Russian Science Foundation, grant No 23-11-00002.

\section*{Appendix: Compact notations}

In this Appendix, we disscuss the notation in which the structure
relations of the Schr\"odinger algebra and its $N=1,2$
supersymmetric extensions take a compact form.

Introducing the notation $H=X_{-1},D=X_0,K=X_{1}$ and
$P_i=Y_{-\frac12,i},C_i=Y_{+\frac12,i}$, the structure relations
(\ref{ShrAl}) can be put into the form
\begin{align}
& [X_n,X_{n'}]=-(n-n')X_{n+n'}, &&
[Y_{-\frac12,i},Y_{\frac12,j}]=m\delta_{ij}, \nn\\
& [X_n,Y_{l,i}]=-(\frac{n}{2}-l)Y_{n+l,i}, &&
[Y_{l,i},M_{jk}^{}]=\delta_{ij}Y_{l,k}-\delta_{ik}Y_{l,j},
\end{align}
where  $n,n'=-1,0,1$, $l=-\frac12,\frac12$ and $i,j,k=1,2,...,d$,
$d$ being a spatial dimension. Above,
the commutators involving the $so(d)$ generators $M_{ij}$ were omitted. Such a notation is convenient for embedding the Schr\"odinger
algebra into the infinite-dimensional Schr\"odinger-Virasoro
algebra (see e.g. \cite{HU05}).

Denoting
$Q=G_{-\frac12},S=G_{\frac12}$ other structure relations of the
$N=1$ Schr\"odinger superalgebra (\ref{ShrAlN1}) read
\begin{align}
& \{G_l,G_{l'}\}=2iX_{l+l'}, && [X_n,G_l]=-(\frac{n}{2}-l)G_{n+l},
\nn\\
& \{G_l,\Lambda_i\}=iY_{l,i}, && [G_l,Y_{l',i}]=-(l-l')\Lambda_i,
\nn\\
& \{\Lambda_i,\Lambda_j\}=im\delta_{ij}, &&
[\Lambda_i,M_{jk}^{}]=\delta_{ij}\Lambda_{k}-\delta_{ik}\Lambda_{j},
\end{align}
where $l,l'=-\frac12,\frac12$.

Similarly, introducing the notation
$Q=G_{-\frac12},S=G_{\frac12}$ and $\bar Q=\bar G_{-\frac12},\bar
S=\bar G_{\frac12}$, the $N=2$ Schr\"odinger superalgebra (\ref{ShrAlN2})
can be compactly written as
\begin{eqnarray}
{\{G_l,\bar G_{l'}\}}&=&2iX_{l+l'}-(l-l')(J+Z),
\nn\\
{\{\bar G_l,\Lambda_i\}}&=&\{G_l,\bar\Lambda_i\}=iY_{l,i},
\nn\\
{\{\Lambda_i,\bar\Lambda_j\}}&=&im\delta_{ij},
\nn\\
{[X_n,G_l]}&=&-(\frac{n}{2}-l)G_{n+l},
\nn\\
{[X_n,\bar G_l]}&=&-(\frac{n}{2}-l)\bar G_{n+l},
\nn\\
{[G_l,Y_{l',i}]}&=&-(l-l')\Lambda_i,
\nn\\
{[G_l,\bar Y_{l',i}]}&=&-(l-l')\bar\Lambda_i,
\nn\\
{[J,G_l]}&=&iG_l,
\nn\\
{[J,\bar G_l]}&=&-i\bar G_l,
\nn\\
{[J,\Lambda_i]}&=&i\Lambda_i,
\nn\\
{[J,\bar\Lambda_i]}&=&-i\bar\Lambda_i,
\nn\\
{[\Lambda_i,M_{jk}^{}]}&=&\delta_{ij}\Lambda_{k}-\delta_{ik}\Lambda_{j},
\nn\\
{[\bar\Lambda_i,M_{jk}^{}]}&=&\delta_{ij}\bar\Lambda_{k}-\delta_{ik}\bar\Lambda_{j}.
\end{eqnarray}

\end{document}